\def\l{$\lambda$}

\def\REF{\par\noindent\hangindent 20pt}
\def\kms{km~s$^{-1}$}

\def\gradip{\hbox{\rlap{\hbox{.}}\raise 5.truept \hbox{$\circ$}}}
\def\mincir{\ \raise-2.truept\hbox{\rlap{\hbox{$\sim$}}\raise5.truept
\hbox{$<$}\ }}

\def\magcir{\ \raise-2.truept\hbox{\rlap{\hbox{$\sim$}}\raise5.truept
\hbox{$>$}\ }}

\def\mincir{\raise -2.truept\hbox{\rlap{\hbox{$\sim$}}\raise5.truept
\hbox{$<$}\ }}
\def\magcir{\raise -2.truept\hbox{\rlap{\hbox{$\sim$}}\raise5.truept
\hbox{$>$}\ }}

\def\a{$\alpha$}
\def\b{$\beta$}

\def\l{$\lambda$}
\def\Msol{$M_\odot$}

\def\ltsima{$\; \buildrel < \over \sim \;$}
\def\simlt{\lower.5ex\hbox{\ltsima}}            
\def\gtsima{$\; \buildrel > \over \sim \;$}
\def\simgt{\lower.5ex\hbox{\gtsima}}            
\documentstyle[12pt]{article}
\begin{document}
\bigskip
\begin{center}
\null
\vspace{4truecm}
\baselineskip=24pt
{\large\bf The peculiar Balmer line profiles of OQ 208$^1$}

\bigskip\bigskip

{\sc P. Marziani$^{2,3}$, J. W. Sulentic$^{2,4}$, M. Calvani$^5$
\\E. Perez$^6$, M. Moles$^7$, M. V. Penston$^8$}

\vspace{1cm}

\baselineskip=20pt

$^2${\small\it Department of Physics \&\ Astronomy, University of Alabama
Tuscaloosa, AL 35487, U.S.A.}

\bigskip

$^3${\small\it International School for Advanced Studies,
Strada Costiera 11, I--34014 Trieste,  Italy}

\bigskip

$^5${\small\it Osservatorio Astronomico di Padova, Vicolo Osservatorio 5,
I--35122 Padova, Italy}

\bigskip

$^6${\small\it Instituto de Astrofisica de Canarias, E--38200 La Laguna,
Tenerife, Spain}

\bigskip

$^7${\small\it Instituto de Astrofisica de Andalucia, Apartado 2144,
Granada, Spain}

\bigskip

$^8${\small\it Royal Greenwich Observatory, Madingley Road,
Cambridge CB3 0EZ, United Kingdom}

\bigskip
\vspace{1.cm}
{}~~~~~~~~~ Received:~~~~~~~~~~~~~~~~~~~~~~~~~~~~~~~
Accepted: ~~~~~~~~~~~~~~~~~~~~~~~~~~~~
\end{center}
\vspace{1truecm}

\leftline{{\bf Running Title:} Balmer Line Profiles of OQ 208}

\vspace{1truecm}
\leftline{\small $^1$ Based in part on observations collected at ESO, La Silla}
\medskip

\leftline{\small $^4$ Visiting Astronomer, Kitt Peak National Observatory}
\newpage
\baselineskip=24pt 
\begin{flushleft}
\parindent=16pt
\newpage
\medskip\medskip\medskip
\leftline {\Large \bf ~~~~Abstract}
\bigskip\bigskip
\baselineskip=24pt

We present spectrophotometric observations of the Broad Line Radio Galaxy
OQ 208 ($\equiv$ Mrk 668 $\equiv$\ 14040+286) obtained between 1985 and
1991. We show that the Balmer line fluxes and profile shapes undergo
remarkable changes. The ratio of intensities between the broad and narrow
components of H\b\ increased monotonically from $\sim 15$\ in 1985 to $\sim
40$\ in 1991. The peak of the broad components of H\b\ and H\a\ were known
to be strongly displaced to the red. We have discovered a correlation
between  the amplitude of the broad peak displacement and the luminosity of
H\b, in the sense that {\it the displacement is larger when the line
luminosity is higher}. The line profiles and the correlation are analyzed
in light of several competing models for the Broad Line Region (BLR). We
conclude that the observations are not compatible with  either  a binary
BLR model or one involving ballistic acceleration of the line emitting gas.

Radiative acceleration of a system of outflowing clouds readily explains
the correlation between line shift and luminosity as well as the peculiar
line profiles. Self-absorption in the Balmer lines or dust absorption in
the non illuminated face of the clouds is necessary to explain the
asymmetry associated with the red peak. Thus it seems that most or all of
the Balmer emission originates from the inward face of the clouds.
Theoretical line profiles computed  for  clouds accelerated by radiation
pressure in various geometrical configurations suggest that the observed
H\b\ profile is best fit assuming the contribution of an ensemble
which might be spherical or confined in a thick disk
in addition to a component emitted in a thin shell contained in a
cone of half opening angle $\sim$ 12$^\circ$ seen along its axis. Another
composite model for the BLR, in which the line is emitted partly by (1) a
relativistic disk seen nearly pole on and (2) infalling matter is not
favored -- but not excluded -- on the basis of the present data.

\bigskip
{\em Subject Headings:} Galaxies: active -- Galaxies: individual (OQ
208$\equiv$ 14040+286) -- Galaxies: kinematics and dynamics -- Galaxies:
Seyfert -- Line: profiles

\section{Introduction}

Balmer line profiles in AGN are presumed to reflect geometric structure and
kinematics in the broad line region (BLR). These profiles show a confusing
array of both red and blue shifts as well as asymmetries (Sulentic 1989). The
importance of the objects showing peculiar profiles has received increasing
recognition in the past few years. It is thought that these objects can
offer a unique insight into the structure of the BLR. Objects with the
largest line shifts or asymmetries might represent, for instance, preferred
source orientations to our line of sight.  Very regular profiles (often
with logarithmic shape) can be successfully modelled employing a variety of
velocity fields (see Mathews \&\ Capriotti 1986 for a review).
Unfortunately this ``degeneracy'' prevents us from obtaining unambiguous
evidence about the kinematics and geometry of the BLR. It is,  for example,
unclear whether the predominant motion in the BLR is rotational or radial
and, if radial,  whether it is predominantly in-- or out--wardly directed.
The broadest and/or most irregular line profiles in Seyfert and Broad Line
Radio Galaxies (BLRG) might, on the contrary, provide a direct link to the
structure of the BLR (see e.g., Penston, 1991; Robinson, Perez \&\ Binette
1990). This is especially true if the profile shape is variable.

Rotation has been proposed as the dominant broadening mechanism for very
broad or peculiar  (sometimes ``double peaked'') Balmer line profiles
(Wills \&\ Wills, 1986; Wills \&\ Browne, 1986). Some profiles have been
fitted using models for a relativistic accretion disk (e.g., Arp 102B:
Chen, Halpern \&\ Filippenko, 1989; Chen \&\ Halpern, 1989; 3C332, Halpern,
1990: 3C 390.3, Perez et al. 1988). Nevertheless, simple disk models are not
able to reproduce the majority of the observed line shifts and asymmetries
in regular profiles
(Sulentic et al. 1990), and the shape of peculiar profiles like those of Akn
120, IC 4329A and others (Marziani, Calvani, \&\ Sulentic 1992).
Refinements of the models (including for instance {\em hot spots}, as in
the case of 3C390.3; Veilleux \&\ Zheng, 1991) can reduce the disagreement
between the expectation for disk emission and the observed profiles. Much
theoretical work in this direction remains to be done.

Spectroscopic data is being collected with growing attention devoted to
peculiar objects but, with the possible exception of Arp 102B, the debate
over the structure of the BLR is still open to many competing alternatives.
The line profile peculiarities have been interpreted within the context of
several models besides the accretion disk scenario including: a) binary
black holes (Gaskell, 1983b), b) transient light--echoes (e.g., Penston
1991 and references therein) and c) a system of clouds in bipolar outflow
(Zheng, Binette \&\ Sulentic, 1990; Zheng, Veilleux \&\ Grandi, 1991).
Furthermore it has been pointed out (Marziani, et al. 1992)
that it is not necessary to invoke {\em ad hoc} exotic mechanisms to
explain the peculiarities of the profiles, since particular geometrical
and/or viewing conditions may account for them. This can also be true under
the assumption that the velocity field is basically the same in objects
with irregular and regular line profiles.

OQ 208 ($\equiv$ Mrk 668 $\equiv$ 1404+28) attracted our attention because
its optical spectrum shows one of the largest peak displacements yet
observed. It also shows some features more typical of Seyfert 1 or BLRG
(Blake et al. 1970; Burbidge \&\ Strittmatter 1972). This object has been
classified as a BLRG because it exhibits rather strong radio emission from
a central compact source. Osterbrock \&\ Cohen (1979, hereafter OC79) made
the first detailed study of the optical spectrum and discovered notable
peculiarities in the broad Balmer line profiles. They pointed out that the
peaks of the broad H\a\ and H\b\ profiles were strongly redshifted (by
about 2600 km/s) with respect to the narrow components of the same lines.

OC79 suggested that a combination of radial motion and obscuration, or self
absorption, could account for the observed profiles. Although OQ 208 has
been extensively monitored at radio wavelengths, even on a daily basis
since it is a candidate flux calibrator (Waltman et al. 1991, and
references therein),  little attention has been paid to its optical
spectrum in the past 15 years. Gaskell (1983a) showed an H\b\ profile
obtained in 1982, and interpreted it within the context of a binary black
hole model. OQ208 is unresolved at VLA resolutions but the radio core has
been mapped recently using VLBI. Two components have been revealed at 8.4
GHz which are separated by $\sim$\ 1.3 mas. Position angle changes
from the highest to the lowest observed frequency suggest that the components
could
be the signature of a highly curved jet observed at different resolutions
(Charlot 1990).

In this paper we present spectrophotometric observations of OQ 208 in the
time period between March 1985 and  February 1991. In \S\ 2 we describe the
observations and reduction procedures.  The results on fluxes and line
profiles are presented in \S\ 3. We find that the H\b\ flux varied
strongly over the period of observation. The peculiar profiles of the
Balmer lines are variable,  and their variations allow  us to place some
unambiguous constraints on the BLR structure. In \S\ 4 we restrict the
choices of possible models able to explain the observed properties of OQ
208 in a simple way. We discuss the advantages and the difficulties of
models involving line emission from: a) a binary BLR, b) an accretion disk
c) an infalling component or a hot spot (\S\ 5), and d) outflow in
the BLR (\S \ 6). We suggest -- also on the basis of the model profile
presented in \S\ 6.2 -- that the most plausible scenario to explain the
Balmer line emission involves a system of clouds accelerated by radiation
pressure in a biconical configuration plus an additional contribution from a
cloud component that is either spherically symmetric or confined within a
thick cylinder. In \S\ 7 we describe additional observations that could test
the model proposed here.

\section{Observations and Data Reduction}

\subsection{Observations}

Spectrophotometric observations of OQ 208 were carried out during the
period 1985 to 1989 with the 2.5 m Isaac Newton Telescope (INT) at La
Palma. The telescope was equipped with an intermediate dispersion
spectrograph and an IPCS detector. The spectra were collected with a slit
width of $\approx$ 1.5 arcsec. An additional, short exposure with a slit
width of 5.0 arcsec was  obtained after each exposure in order to ensure an
accurate spectrophotometric calibration. The slit was oriented at
parallactic position angles to minimize wavelength dependent losses. The
use of a 300 l/mm grating yielded a dispersion of $\approx$ 2 \AA/pixel.

An additional spectrum of OQ 208 was obtained in April 1990 with the 1.52 m
telescope at the European Southern Observatory on La Silla. The telescope
was equipped with a Boller \&\ Chivens spectrograph and an RCA high
resolution CCD detector (pixel size 15$\mu m\times$15$\mu m$). The slit
width was $\approx $ 2.0 arcsec at the focal plane of the telescope.
The seeing was estimated to be $\simlt$ 1 arcsec during the observation.. A
1200 l/mm grating allowed a dispersion of 60 \AA/mm. OQ 208 was
also observed at  Kitt Peak National Observatory during February 1991 with
the Gold  Camera attached to the Cassegrain focus of the 2.1 m telescope.
An 800 $\times$ 800 pixel TI chip and a 600 l/mm grating were used yielding
a dispersion of 1.3 \AA/pixel. The slit width was 2.5 arcsec and the
angular scale at the focal plane of the spectrograph was $\sim$0.8
arcsec/pixel. The seeing was estimated to be 1.5 arcsec.
The Journal of Observations given in Table 1 lists the date when each
spectrum was obtained along with the exposure time, spectral dispersion  and
the
wavelength range covered.

\subsection{Data Reduction and Preliminary Analysis}

A detailed report of the data reduction procedure for the spectra taken at
the INT is given in Perez (1987), while for the spectrum taken at ESO,
detailed information can be found in  Marziani (1991) and Marziani, et al.
(1992). Only the basic steps of the reduction procedure are
reported here.

The bias level was subtracted from all spectra, which were then divided by
a flat field.  Wavelength calibration was obtained from comparison spectra
taken immediately after (before and after the KPNO observation) the
spectrum of the  galaxy. Standard IRAF and Vista procedures were followed
to convert the spectra to a linear wavelength scale. The rms was $\simlt$
0.1 \AA\ for the ESO and Kitt Peak spectra. The instrumental resolution was
estimated measuring the FWHM of faint lines of the comparison
spectra. The values are approximately 2 \AA, 3.2 \AA,  4 \AA\ for the ESO,
KPNO and INT observations
respectively. Flux calibration was obtained from at least
two/three nightly observations of spectrophotometric standard stars. We
estimate, from a comparison of the [OIII]\l\l 4959,5007 flux measurements
in different spectra, that the uncertainties in the flux calibration are
$\approx \pm 25 $ \%\ (2$\sigma$).

The spectra were scaled to the average of the measured [OIII]\l\l4959,5007
fluxes, in order to compare the relative fluxes of the broad lines at
different epochs of observations. Although the spectrum obtained at KPNO
was calibrated using standard stars observed with a narrow (2.5 arcsec) slit,
the results for the narrow lines are in good agreement with the average of
the INT observations. We are therefore confident that the calibration is
adequate.

The scaling procedure is somewhat arbitrary for the spectrum taken at ESO
because the
[OIII]\l\l4959,5007 lines were outside of the observed spectral range. The
flux level in the ESO spectrum is discordant with respect to the others by
a large factor $\sim$ 2.5. The discrepancy is too large for us to use it
with confidence. It has been used  for the analysis of the H\a\
profile shape since it provides the highest S/N ratio in the H$\alpha$\
region. No line fluxes are reported for the ESO spectrum in Table 2.

Two--dimensional sections of the CCD frames obtained at ESO and at KPNO
were wavelength and flux calibrated, to allow for study of possible
extended emission in OQ 208.

Uncertainties quoted throughout the paper are at the 2$\sigma$\ level unless
otherwise stated.

\section{Results}

\subsection{Line Identification and Line Fluxes}

The overall optical spectrum of OQ 208 is shown in Figure 1. The
H\b\ and H\a\ regions of four representative spectra are shown in Figure 2
(after normalization to the same [OIII]\l\l 4959,5007 flux) at
a scale which emphasizes changes in the level of the underlying
continuum and variations in the line profiles. The heliocentric radial
velocity of OQ 208 measured from [OIII]\l\l 4959,5007, [NII]\l 6583, and
the narrow components of H\b\ and H\a\ is $v_0 = 22985 \pm$ 15 \kms. The
absolute line fluxes of the most prominent narrow lines (corrected for
galactic reddening) are reported in Table 2. The large shift between the
peaks of the broad and narrow components in H\b\ allows us to
estimate their relative contributions with high accuracy. The H\b$_{NC}$\
flux estimates agree within $15$\ \% (after normalization to their
respective [OIII]\l\l4959,5007 fluxes and excluding the spectrum obtained
in June 1985).
Deblending  of the narrow H\a\ and [NII]\l\l6548,6583 components from the
underlying broad H\a\ profile was not so easy. Any deconvolution procedure
will be highly subjective because the peak of broad H\a\ is shifted to
the red side of [NII]\l 6583. Differences between the broad H\a\ and H\b\
profiles make it difficult to use a suitably scaled  H\b\ profile as a
model template for the broad component of H\a. Nevertheless, while not
matching the {\em total} broad H\a\ profile, we note that the scaled H\b\
profile reproduces quite well the red wing (see Fig. 2e and 2f
and the analysis in \S\ 3.2). This suggests that the shape of broad H\a\
emission underlying H\a$_{NC}$ and [NII]\l6583 is properly taken into
account by such a scaling. We repeated this procedure for all of the
spectra with reasonable S/N and the deblending yielded very similar
results. We find that [NII]\l6583/H\a$_{NC}$ $\approx 0.98 \pm 0.10$, and
$H\alpha_{NC}/H\beta_{NC} \approx 4.1 \pm 0.8$. The internal reddening
estimated from the Balmer decrement is therefore E(B--V) = 0.175. Fluxes of
the fainter narrow lines reported in Table 2 are believed to be accurate
within $\pm 50$ \%. The large uncertainty is a result of the low S/N ratio
in the H$\alpha$\ spectral region.

Fluxes of the prominent broad lines and FeII blends are reported in Table 3
along with values for the continuum at \l 4800 \AA. Column headings in
Table 3 indicate the dates of observation. The spectrum of OQ 208 is
atypical for a BLRG or a Seyfert--1 galaxy if one considers the large peak
displacement of the broad Balmer lines. The value of the ratio
[OIII]\l5007/ H\b$_{NC}$\ is however typical for BLRG although rather high
for a Seyfert 1 galaxy (Osterbrock, 1978).  The [OIII]\l\l 4959,5007 lines
show a fainter component displaced 1000 km s$^{-1}$\ to the red.  Heckman,
Miley \&\ Green (1984) using lower S/N data describe this second component
as broad [OIII]\l\l4959,5007. It is doubtful that broad [OIII]\l\l
4959,5007 could significantly affect the red side of H\b\ in OQ 208. The
FeII lines at 4924 and 5018 \AA, which are expected to have broad profile
similar to that of H$\beta_{BC}$, are clearly present  (it is interesting
to note that they fall at the positions expected if their peak displacement
is the same as the red peak of H\b). These lines can mask any weak
broad-line component of [OIII], especially if the FWHM of the broad lines
is larger than 3000 km/s (Crenshaw \&\ Peterson, 1986). The H\b\ profile
resulting from the subtraction of the narrow [OIII] profile and the FeII
emission is rather smooth. We conclude that there is no evidence for broad
[OIII] emission in the red wing of the H\b\ profile.

In the spectrum
obtained at Kitt Peak we found no  difference between the cross--dispersion
Point Spread Function of the continuum and the [OIII]\l\l 4959,5007 lines.
This suggests that the [OIII]\l\l 4959,5007 emission must arise in an
unresolved region within $d \approx$ 1.6 arcsec.

FeII emission on the red side of H\b\ is clearly visible in the Figure 1
spectrum obtained when OQ 208 was near the bright stage. However, FeII
emission on the blue side of H$\beta$\ falls  below the noise level in many
spectra (we will henceforth refer to the FeII emission  \l 4460-4700 \AA\
and \l 5190-5320 \AA\  as the 4570 and \l 5250 blends respectively). In
Table 2 we give an estimate for the total flux in the FeII blends and an
upper limit for the flux when the \l 4570 emission falls below the
5$\sigma$ continuum noise level. The  ratio of the total FeII emission to
H\b\ is not uncommon among Seyfert 1 galaxies and is also not unusual for a
compact--core dominated radio galaxy (Joly, 1991; Jackson \&\ Browne,
1991).

The ratio between the FeII\l 4570 and \l 5250 blends is in the range
0.14--0.33. This is unusual for a Seyfert 1 galaxy. The average value of
the ratio FeII\l 4570/FeII\l 5250\ for the 18 Seyfert galaxies  measured by
Phillips (1978) is $\sim 1.05 \pm 0.25$.   Furthermore, assuming a flux
ratio between the radio core and lobes equal to  $R = 2.3$\ (Perez, 1987),
the ratio  FeII\l 4570/[OIII]\l 5007 is low for a compact core radio galaxy
(Jackson \&\ Browne, 1991). Since the ratio FeII\l 5250 /H\b$_{BC}$ is much
more typical, we conclude that emission from FeII\l 4570 is unusually low
in OQ208.  Only Mkn 231, among the objects studied by Phillips (1978),
shows a \l4750/\l5250 ratio ($\approx$ 0.4) similar to OQ208. This object
is one of the three strongest FeII emitters known (see e.g., Sulentic et
al. 1990).  Internal reddening in Mkn 231 is probably responsible for a
steep decrease in the spectrum toward shorter wavelengths. The steep
continuum observed in the spectrum of OQ 208 (Fig. 1) suggests that
internal reddening may play a role in this object as well. The low FeII
ratio could also related to excitation conditions in the FeII emitting
zone. In fact, faint emission is suspected in the region 6000--6500 \AA,
where contributions from multiplets 199 \&\ 200 are expected if the main
energy input is provided by photoionization. This result is somewhat
peripheral to the present investigation. It requires confirmation by
further high S/N observations in the red spectral range.

\subsection{Line Variability \&\ Line Profiles}

The flux of H\b\ underwent remarkable changes by a factor $\simlt 2-2.5$ in
the period 1985 and 1991. This is well demonstrated from variations in the
ratio I(H\b)$_{BC}$/ I(H\b)$_{NC}$. The value of this ratio increased
monotonically from  $\sim 15 $  in March 8, 1985 to $\sim 40$ \ in Feb. 19,
1991. We also measured this ratio on an enlarged print of the 1978 spectrum
taken by OC79. We found that I(H\b)$_{BC}$/ I(H\b)$_{NC}$ $\sim 40 - 50$,
somewhat larger than the value obtained in the 1991 spectrum. Since there
is no doubt that this ratio was much smaller in 1985, the line luminosity
must have faded between 1977 and 1985 followed by an increase between 1985
and 1991. The variations suggest that the Balmer lines may be described as
passing through a {\em high} and {\em low} luminosity phase. The H\b\ flux
from the 1991 observation is the largest among our spectra, and we consider
this profile as representative of the high phase. Evidence for recurring
high and low phases was also found for Akn 120 and IC 4329A (Marziani, et
al. 1992). No strong variations in the emission line profiles were observed
in these two galaxies and the luminosity changes suggested a shorter
time--scale.

The difference between the spectrum obtained on Feb. 19, 1991 (high phase)
and the spectrum obtained from the weighted (over the S/N ratio) average of
the spectra taken in 1985 (low phase) is shown in Fig 3. The major changes
of the H\b\ profile can be described as an increase on the red side of the
peak and a remarkable increase of the blue wing. No obvious change occurred
in the red wing, a result which should be confirmed due to the [OIII]\l\l
4959,5007  emission in this part of the profile.

The principal result of this investigation involves the discovery of a {\it
correlation} between the H\b\ line luminosity and the radial velocity of
the red peak. The center of the flat--topped red peak observed in the 1991
spectrum  is shifted with respect to the narrow component of H\b\ by
$\Delta v_r \approx$ 2600 km $s^{-1}$ (see Fig. 2e). The spectrum published
by OC79 shows a similar value for the velocity displacement. In order to
verify that a change in the peak displacement occurred, one can consider
that an upper limit to the displacement of the peak in 1985 (see Figure
2a), when the line luminosity was much lower  was $\Delta v_r\approx$ 1500
km/s. A spectrum taken in April 1985 with the 3.0 m Shane telescope at Lick
Observatory, and kindly provided to us by R. Cohen, confirms our estimates
for the line ratio and the peak shift.

The definition of a peak or centroid radial velocity is by itself subject
to ambiguity and to many uncertainties. The situation is made even worse
for OQ 208, since the broad H\b\ profile at certain epochs is flat--topped,
and since the S/N in some spectra is low. We first measured the
``centroid'' by eye using the peak wavelength of the broad component or the
middle of the flat top. In order to analyze the peak displacement in a more
quantitative fashion, we also considered the centroid of the red peak weighted
on the third power of the intensity for each wavelength interval (see Fig.
4a, where the intensity ratio between the narrow and broad component of
H\b\ is plotted versus line shift).

Analysis of Figure 4a suggests that the peak shift and line strength are
correlated.  The limited number of observations, especially at different
stages in the variability, prevent us from describing the functional form
of the correlation. We note that variations in flux and shift occuring in
1985 and 1986, as well as in 1988 and 1989, are not much larger than the
observational uncertainties. Significative changes appear to occur only
between 1986 and 1988 and between 1989 and 1991.
Data appear to be correlated irrespective of the definition of the
centroid. However, the amplitude and
shift change somewhat depending upon the choice of the window over which
the centroid is computed.

Figure 4b shows the H\b\ peak displacement plotted against the normalized
intensity measured between H\b$_{NC}$ and the [OIII]\l 5007 line. This
is equivalent to measuring the line flux emitted
within a radial velocity interval equal to $\approx \pm$ 0.13 HWZI. This will,
consequently, exclude a significant contribution from the line wings and
will emphasize the variation of the red peak. We favor this procedure
because the velocity frame for the flux measurements should be set at the
radial velocity of the redshifted peak. We just avoid that the origin will
be different at each epoch because the redshift of the peak is changing.
The data are
correlated with approximately the same degree of confidence.
This suggests that the red peak is probably related to an independent
 component which varies more strongly than the rest of the line. This
interpretation is supported: (1) by the shape and lower redshift of the
broader part of the H\b\ profile and (2) because the percentage of variation
in the broad component is smaller than for the red peak. Figure 4c shows
the H\b\ shift versus the continuum flux at 4800 \AA. Although the estimate
of the continuum flux is sensitive to the uncertainties in the
spectrophotometric calibration, the data show a correlation.


Analysis of the H$\alpha$\ profile is made difficult by the large spread in
S/N  among the red sensitive spectra. The strength and shape of the blue
wing on the H\a\ profile has probably varied in the spectra taken at
different epochs. The shape of the blue wing in H\a\ and H\b\ was similar
when the line luminosity was near minimum (1985--1986). The blue wing of
H\a\ appears to have increased in strength relative to the red one in the
spectra taken after 1988. This result is supported by the change in the
blue-wing of H\b\ shown in Fig. 3, but it remains  uncertain to some
extent since the H\a\  region in the INT
spectra was not corrected for B-band absorption. If we compare the
profiles of H\b\ (Feb. 19, 1991, Fig 1e) and H\a\ (Apr. 4, 1990, Fig. 1f)
we note
that the shape of the red wings in both lines are very similar except that the
blue side of H\b\  lacks the hump clearly visible in H\a. This suggests at
first that the Balmer decrement is steeper in the gas emitting the blue
part of the line.

The blue wing of H\a\ takes the form of a peak in the ESO (1990) spectrum.
Note that this blue peak is also present in OC79. This feature is probably
real, although it is not clearly seen in the spectra obtained at the INT.
We note however that the blue side of the H\a\ profile is heavily
contaminated by the telluric B--band absorption. The ESO spectrum was
corrected for B--band absorption by using observations of a standard star
observed at a similar zenith distance immediately after OQ 208. This was
not possible for the other spectra.

\section{Discussion}

The observation of a large redshift for the broad Balmer lines suggests an
AB,R classification (i.e., blue asymmetric profile with redshifted peak)
for OQ208 according to the scheme proposed by Sulentic (1989). This appears
to be a rare class. On the contrary, AR, B profiles appear to be quite
common.  Of the 61 objects included in Sulentic (1989) only I Zw 1 was
assigned type AB,R. The redward displacement of the broad component in I Zw
1 ($\Delta$V $ \approx$ 640 km/s) is much less than for OQ208. I Zw 1 also
shows a much narrower profile that is contaminated with strong FeII
emission. The two objects show a similar displacement expressed in units of
their respective profile FWHM. The degree  of profile asymmetry in I Zw 1
is not nearly as pronounced (asymmetry index of 0.07 compared to 0.2 for
OQ208). If the AB,R classification reflects similar geometry in these two
objects, then the FWHM of the Balmer lines is somewhat independent of
geometry considerations.

A number of objects with peculiar line profiles have been studied by
various authors in the past few years. A partial list includes 3C390.3,
3C382, Arp 102B, 3C332, OX 169, Akn 120, IC 4329A (see Marziani 1991 for
spectra and references). The objects listed above can be divided into two
subgroups: I) OX 169, Akn 120 \&\ IC 4329A show a redshifted peak in
addition to a peak located at approximately the same redshift as the
underlying galaxy ( a small blueshift of $\approx$ 100 -- 300 km s$^{-1}$
is suspected for this ``zero'' peak); II)  Arp 102B, 3C332, 3C390.3, 3C382
show two prominent peaks {\em roughly} symmetrically displaced with respect
to the systemic radial velocity. Apparent differences between objects of
the same subgroup exist, particularly as far as variations in the profiles
are concerned. Prototype objects with a double peaked profile,  Arp 102B
(Chen \&\ Halpern, 1989, and 3C332 Halpern, 1990) show rather stable
emission features. Although the possibility of variations have been
suggested (Halpern \&\ Filippenko, 1988; Miller \& Peterson, 1990), they
appear to be small and not comparable with the changes observed in the H\b\
profiles of 3C390.3 and 3C382 (Perez 1987; Perez et al. 1988). OQ 208 is
somewhat unique in that it shows only a red peak, although its H$\alpha$\
profile resembles the profile observed in 3C390.3. The similarity with
3C390.3  (cf. Figures 5.13 of Perez 1987) holds only at certain epochs: 1)
variations of the line profiles in 3C390.3 are of larger amplitude and
occur on a shorter timescale than variations in OQ 208 (e.g., Veilleux \&\
Zheng, 1991) and 2) unlike OQ 208, the spectra  of 3C390.3  show a more
prominent blue peak at certain epochs. A 900 km/s change in the {\it blue}
peak displacement has also been observed.  Veilleux \&\ Zheng (1991) and
Zheng, Veilleux \&\ Grandi (1991) suggest that the variations in strength
and peak displacement may be due to the changing position of a hot spot on
the disk.  We will discuss later the relevance of this model to the case of
OQ 208.

We propose to model the BLR structure of OQ 208 taking advantage of
the main observational facts established in this investigation:

\begin{enumerate}
\item the flux of the Balmer lines varied by a factor of 2--3 between 1985
and 1991;
\item the red displaced broad component peak remains the most prominent feature
in the H\a\ and H\b\ profiles at all epochs of observation;
\item  the radial velocity of the red peak in H\b\ is correlated with the
line luminosity.
\end{enumerate}

OC79 proposed that the large redshift of the Balmer line peaks in OQ208 was
gravitational in origin. The addition of gravitational and transverse
redshift can lead to $\Delta z \approx \frac{3}{2} \frac{R_{g}}{R}$, where
R is the distance of the emitting gas from the central black hole, and
$R_g = 2GM_{BH}/c^2 \approx 2.95\times 10^{12} M_{BH,7}$\ cm
is the gravitational radius of the black hole.
$M_{BH,7}$ the black hole mass is units of 10$^7$\ $M_\odot$.
The redshift is
however too large to be entirely due to the gravitational field of the
black hole. The shift $\Delta v_r \approx 2600$ \kms would imply that the
bulk emission of the line takes place at $R \simlt 150 R_g \approx
4.38\times 10^{14} M_{BH,7}$\ cm, an implausibly small distance between the
BLR and black hole. It would be a factor 5--7 below estimates from
photoionization calculations, and smaller than the inner radius of the BLR
estimated from cross--correlation analysis, even assuming anisotropic line
emission (e.g., Rees, Netzer \&\ Ferland, 1989; Ferland et al. 1992).

The  ``hot spot'' interpretation also faces serious difficulties. If we
accept the nearly face-on orientation of the disk implied by the R
parameter, it is not clear how to explain a lower luminosity when the
displacement is smaller. Obscuration of the rotating spot by the disk
itself could in principle explain this result but obscuration plausibly
occurs only if the disk is seen nearly edge-on. In addition, changes in
radial velocity due to radial drift of the hot spot are expected to occur
on timescales which are much longer than the Keplerian timescale for a
geometrically thin disk.

It seems more realistic to ascribe  the large peak redshift to bulk motions
of the BLR gas. The observation of a shifted broad component in H\a\ and
H\b\ points toward a predominance of radial motion over rotational or
random motions in a virialized ensemble of clouds. The unusually large
value of the shift probably points toward an extremum in the viewing angle
of the AGN: i.e., the bulk emitting gas should move in  a direction along
(or at least not too far from) the line of sight.

Ideas regarding orientation indicators for the central engine of AGN
deserve some discussion at this point. Orr \&\ Browne (1982) suggested that
the appearance of a radio source is governed by the viewing angle of the
observer with respect to the radio axis: if the radio galaxy is observed
pole-on, we see a compact core dominated source and if the source is
observed edge-on, we see a lobe dominated source. A relevant implication of
this view is that the radio morphology can be used as an ``aspect
indicator'' and that it provides an estimate of the inclination at which
the radio axis  is observed with respect to the line of sight. This
procedure also gives the orientation of the accretion disk assuming that
the plane of the disk is perpendicular to the radio axis. In our case, the
compact radio source associated with OQ 208 would imply that the disk is
seen nearly face on. The parameter R defined by Orr \&\ Browne (1982) is
thought to be the most reliable inclination estimator available and in the
following we will consider with some confidence the assumption of an
accretion disk oriented face on in OQ 208.

\section{Non--Radial Considerations}

A coarse trend  between H\b\ FWHM and the R parameter for BLRG (objects
with the broadest Balmer line profiles tend to be lobe dominated sources,
Wills \&\ Wills 1986) suggests that the Balmer lines are emitted in a plane
perpendicular to the radio axis and, possibly, by the accretion disk. This
trend is not  well established since, for each value of R (and hence for
each disk inclination) there is a large spread in FWHM.

Apart from the heuristic considerations outlined above, the observation of
a red peak much stronger than a blue one disagrees with the predictions of
simple relativistic disk models. This makes it very difficult to argue that
the {\em entire} Balmer line emission of OQ 208 originates from a Keplerian
disk. Relativistic accretion disk models produce line profiles with an
enhanced blue peak (due to Doppler boosting) and a slightly redshifted line
base (due to the combined effects of gravitational and transverse redshift;
Mathews, 1982; Chen \&\ Halpern, 1989).  We have noted that the blue wing
of H\a\ in OQ208 can be much stronger than the blue wing of H\b\ and that
the blue wing of H\a\ has the form of a broad hump in the 1990 spectrum
(and in OC79). If the blue hump is assumed real and is interpreted as
arising from a radiating disk, a relativistic disk model profile can be
made to fit the broad component of H$\beta$ (with power--law emissivity of
index $q=3.2, R_{in} = 200 R_g, R_{out} = 1250 R_g$, inclination $i
=37^\circ$, with $i=0^\circ$ corresponding to a disk oriented  face--on).
The addition of a  redshifted component, needed to explain the red peak,
provides a satisfactory reproduction of the broad component of H\a\ as
observed in 1990. A disk model profile able to reproduce the form of the
blue H\a\ hump suggests that the disk emission would contribute about
$\approx 2/3$ of the H$\alpha_{BC}$\ flux. If we assign most of the red
peak to an independent component and accept the reality of the blue peak in
H\a\, the  profile observed in 1990 can be made consistent with an
accretion disk model.

Another possibility to consider before discussing radial motion is that the
large displacement of the red peak in H\b\ is due to the orbital motion of
a binary black hole. Two constraints are relevant to this model: a) spectra
exist for the period 1985-91 and b) a lower limit to the variation in peak
radial velocity is $\Delta v_r \approx 1000$ km/s. This suggests that the
phase change of the binary should be: $\Delta \Phi = arccos (1 -  \Delta
v_r/v_r) \approx 53^\circ$ which implies that the  period is $\approx$ 37
yr, and that the mass of the binary is $M_{BH} \simgt \frac{1}{338} P
V^3_{r,1000} \approx 1.7\times 10^8 M_\odot$. There are three observational
points listed near the beginning of this section that argue strongly against
this result. The second point makes it necessary to assume that the center
of mass of the binary has a different radial velocity from the reference
frame of the underlying galaxy. The third point is perhaps the strongest
argument against the binary idea. In the binary model, changes in the line
$\Delta v_r$ should only be related to the orbital motion of the binary
black holes, and should not be correlated with changes in the line or
continuum fluxes.

\section{Radial Considerations}

\subsection{General constraints}

The above considerations point towards radial motion as the most likely
cause of the red peak in OQ208. The observed correlations between the peak
redshift and line and continuum fluxes (Fig. 4a, 4b and 4c) indicate that
the velocity field is probably (although not necessarily, as discussed
below) coupled to the continuum luminosity of the object. Furthermore, the
correlation rules out the possibility that the line emitting gas can
acquire its momentum on a time scale short compared to the time scale of
the variation; i.e., the motions of the clouds can not be ballistic.

We must next consider whether the  motion of the clouds is directed inward
or outward. We can imagine a system of clouds surrounding the accretion
disk. If the disk is oriented pole on, as we have proposed for OQ208, we
might be able to see clouds located on the nearer side of the accretion
disk preferentially. In this case emission from the clouds would only be
redshifted if the clouds were infalling. The clouds moving at the highest
velocity should be located closest to the central source if they are freely
falling in the potential well of the black hole. This would not necessarily
be true if the effect of radiation pressure is taken into account. The
clouds would move at lower velocity when they are closer to the central
source if radiative deceleration overcomes gravity (e.g., k $<$ 0 in Eq.
2 of \S\ 6.2).
In principle, clouds could approach the central source and be pushed
away by radiation pressure. This scenario leads to a prediction that
contradicts our observations (point 3 above): if radiation pressure
decelerates infalling clouds, an increase in the continuum luminosity would
enhance the radiation pressure and the line luminosity, but it would
also lead to a decrease in the velocity $\Delta v_r$ of the infalling gas.

If the clouds are radiation--bounded and the continuum luminosity changes,
we expect a change in emissivity (occurring first for clouds closest to the
continuum source) without any appreciable change in the velocity field.  If
this case is applicable, and radiation pressure is negligible, the gas
moving at higher velocity (located closest to the continuum source) simply
has a {\em stronger} response than the gas moving nearer to the peak
velocity. This touches on a basic problem in the understanding the BLR. It
is unclear whether the structure of the BLR changes appreciably in response
to strong continuum changes. In the following we will assume that the
effects of radiation pressure are likely to be significant.

The case for infall remains appealing also because of the redshifted
secondary component in [OIII]\l\l4959,5007. This feature is suggestive of
infall (see e.g., Rafanelli \&\ Marziani, 1992, for a case in which [OIII]
lines show a strong red--ward asymmetry, and for which infall has been
established). If there is a continuity between the properties of the gas in
the NLR and in the BLR, as several arguments suggest (e.g., Appenzeller \&\
Ostreicher, 1988), both the NLR and BLR could be infalling toward the
central source.

If the gas emitting the Balmer lines is optically thick to the ionizing
continuum, changes in the line fluxes are proportional to changes in the
continuum. Morris \&\ Ward (1988) and Zheng (1991) have shown that the
Balmer lines of several AGN are emitted by optically thick gas with the
possible exception of the far wings, so that it is legitimate to assume
that, at least, the red peak of H\b\ is emitted by optically thick gas.
Recent results by Ferland et al (1992) support this point of view and
suggest that the BLR clouds radiate anisotropically. The correlation
between shift and line intensity suggests that radiation pressure is
dynamically important.

The observation of a red peak stronger than the blue one poses some
difficulties, if we do not rely on optical depth effects for the Balmer
lines.  The stronger peak should be the blue one in this case -- the
contrary of what is observed. Thus, the gas can be accelerated outward by
radiation pressure only if the optical depth is very large in the Balmer
lines or if dust on the back of clouds absorbs the outcoming radiation --
allowing Balmer photon to escape preferentially  from the illuminated face
of the clouds. It has been usual in the past years to ignore optical depth
effects in the Balmer lines when computing profile models. The second level
of hydrogen inside the BLR is overpopulated because of Ly\a\ trapping and, as a
consequence, the optical depth in the Balmer lines is expected to be high.
The escape probability of a Balmer photon from the non--illuminated face
of the cloud should in turn be very low (Ferland et al. 1992).

This motivates us to model the profile of H\b\ (as observed in 1991) under
the assumption that the line is emitted by a system of clouds moving
radially outward under the combined effects  of radiative and gravitational
forces. We assume a biconical geometry (which includes the case of a
spherically symmetric system if the half opening angle of the cone is
$\theta \approx$ 90$^\circ$).

The correlation shown in Figure 4b suggests that the increase in continuum
luminosity produces an enhancement of a radially moving component. The red
peak varies more strongly than the broader base.  The presence  of two
components in H\b\ is supported by the shape of the profile, which shows an
inflection at the base of the red peak. Moreover, the profile difference
between the high and the low phase (Fig. 3) shows a narrow peaked feature
which is responsible for the increase in the centroid shift.

Our interpretation of the variations observed in H\b\ and of the {\em best
fit} to the line profile (described in the next section) have several analogies
with the findings of Ulrich et al. (1985) concerning the appearance of
emission features on the blue and red side of CIV\l 1549. The ``satellite''
lines to CIV\l1549 were interpreted as emission coming from BLR clouds
trapped in a jet. We interpret the second component of H\b\ in basically
the same way. We interpret that lack of a blue ``satellite'' as due to  the
emission anisotropy and the orientation of OQ 208.

\subsection{The model}

The previous discussion of broad line profile shapes, and consideration
of various models,  favors a scenario in which outflowing clouds are
driven by radiation pressure (e.g., Blumenthal \&\ Mathews, 1975). This
model is not devoid of theoretical difficulties, since the existence of
clouds requires a hot, less dense confining medium with which the clouds
should be in pressure equilibrium. Drag forces and hydrodynamical
instabilities could lead to cloud disruption in a time shorter than the
dynamical timescale (Mathews \&\ Ferland, 1987; Mathews \&\ Veilleux,
1989). The confining medium should give rise to some signature in the
X--ray spectrum. This is not observed in the spectra of Seyfert galaxies
and quasars (Mathews \&\ Ferland, 1987, Osterbrock 1991).

The equations of motion for clouds moving under the effect of radiative and
gravitational acceleration was studied by Blumenthal \&\ Mathews (1975).
In the following discussion we assume that the acceleration due to
optical and infrared radiation is negligible compared to that from the ionizing
radiation. The question of whether the clouds are outflowing or infalling
depends upon the acceleration parameter, namely

\begin{equation}
k = \frac{ A_c \int_{\nu_0}^{\infty} L_\nu d\nu}{4 \pi c M_c} - GM_{BH}
\end{equation}

where $A_c$ is the area of the cloud exposed to the ionizing continuum,
$\nu_0$ is the Rydberg frequency, $L_\nu$\ is the specific luminosity of
the ionizing continuum, $M_c$ is the mass of a single cloud, and $M_{BH}$\
is the black hole mass. The expression for k can be rewritten in a more
convenient form:

\begin{equation}
k = \frac{\int_{\nu_0}^{\infty} L_\nu d\nu}{4 \pi c \mu N_c} - GM_{BH}
\end{equation}

where $N_c$\ is the column density, and $\mu$\ is the mean molecular
weight. Holding $M_{BH}$\ as a free parameter, the latter equation contains
quantities which can (in principle)  be estimated from observations in a
straightforward manner. We assume in addition that $M_c$\ and $N_c$\ ($\sim
7 \times 10^{22}$\ cm$^{-2}$) do not change during the cloud's motion. This
might not be a realistic assumption if drag effects alter the shape as well
as the mass of a cloud (Mathews \&\ Veilleux 1989).
Unfortunately there are no direct observations of the ionizing continuum in
OQ 208 that are available to us. We are therefore forced to a somewhat
indirect estimate of the ionizing luminosity. We deduced the number of
ionizing photons emitted by the continuum source from the reddening
corrected H\b\ luminosity assuming a covering factor of $f_c \approx $\
0.17, which is the average value for the Seyfert galaxies studied by
Padovani \&\ Rafanelli (1988). We then calculated the ionizing luminosity
assuming the spectral shape of the ionizing continuum has the form
considered by Netzer (1990). We obtained $L_{ion} \approx$
5.9$\times$10$^{44}$ h$^{-2}$ erg s$^{-1}$\ ($H_0 = 100\times h$\ \kms
Mpc$^{-1}$). It follows that outflow is possible ($k > 0$) if the black
hole mass is $M_{BH} \simlt 2.3\times10^7$ h$^{-2}$ \Msol. A mass of $10^7$
\Msol\ is plausible for OQ 208 since its ionizing luminosity gives a value
for the Eddington ratio that is within the limits set by the estimates of
Padovani \&\ Rafanelli (1988).

The equation of motion yields a velocity field of the form:

\begin{equation}
u(r) = \sqrt{A - \frac{2k}{r}}
\end{equation}

in the case of outflow (k $> $ 0), where $A = 2k/R_{1}$, and $R_{1}$\ is
the inner radius of the BLR.  This function is
typical of outflowing winds (e.g., Wallerstein et al. 1984).

The integral equation relating the line profile $P(\lambda)$\ to the
emissivity, the cloud density and the velocity field can be written as:

\begin{equation} \displaystyle
P(\lambda) = 2 \pi \int_{R_{min}}^{R_{max}}
\int_{-\Theta_0}^{\Theta_0}  r^2 \sin \theta d\theta
dr j_c(r) n_c(r) \delta [\lambda - \lambda_0 (1 + \frac{u(r) \xi}{c})]
\end{equation}

where the opening angle of the cones is $2 \Theta_0$, and the BLR is
assumed to extend from $R_{min}$\ to $R_{max}$, and $\xi = \cos \theta $,
is the cosine of the angle between the line of sight and the velocity
vector of each cloud. In principle, the number density of clouds $n_c(r)$\
can be computed from the continuity condition $n_c u(r) r^2 = const$. Since
this law  probably breaks down at the inner edge of the BLR where the
clouds are assumed to form, we considered also power-law  functions for
$n_c(r)$, namely $n_c(r) = n_0 (r/R_{min})^{-n}$. Template profiles have
been computed assuming that the  cloud emissivity $j_c(r)$ is again a
gaussian or a power-law ($j_c(r) = (r/R_{min})^{-m}$, with $m = 1,2,0,-1$).
A grid of profiles was computed for different values of $\Theta_0$ and
inclination of the cone axis with respect to the line of sight ($i$). The
inner edge of the BLR was estimated at $R_{min} \approx 10^4 R_g \approx
2.95\times $10$^{16} M_{BH,7} $\ M$_\odot$ ~cm, while the outer
radius was set at $R_{max} = 10^5 R_g$. The parameter $k$\ was computed
assuming $M_{BH} = 10^7$ \Msol, $N_c = 7\times 10^{22} cm^{-2}$. We further
assume that the clouds are accelerated from rest.

We remark that the two main conditions which must be satisfied are that:
(1) the peak of the profile is displaced and (2) the peak is nearly flat.
In order to achieve a global shift of the line we assume that the
approaching half of the flow contributes little or no emission. If the
emission of the Balmer lines is intrinsically anisotropic (see e.g.,
Ferland et al. 1992 or Zheng, Binette \&\ Sulentic, 1990, where the effect
of anisotropy is maximized since the opening angle of the cone is assumed
small and the line of sight close to the cone axis), there is little or no
need to introduce any additional source of obscuration. The effect of the
anisotropy is assumed to be proportional to $\sin (\theta/2)$ and
$\sin(\theta/2 + \pi/2)$ for clouds located on the near and far side
of the continuum source respectively.

Single component models (including anisotropy) produce profiles which are
roughly similar to the observed H\b\ profile of OQ 208. We were unable
however to satisfactorily fit the details of the profile.  It is
interesting to note that clouds in a curved jet would also give rise to
profiles with shift and asymmetry similar to those observed in OQ 208
(Wallerstein et al. 1984). We did not attempt a fit in this case.

We modelled the lines as made up of a component emitting the broad wings
and another emitting  the red peak (see Fig. 5). The shape of the broad
wings suggests that the line base is emitted by a spherical ensemble of
clouds surrounding the central engine. This result is not strongly
dependent on the  velocity field or emissivity. We note that the inclusion
of anisotropic emission lead us to a very good reproduction of the
asymmetry and line shape of H\b. Satisfactory fits can be obtained if the
clouds are confined in a sphere (as is the case for the fit shown in Fig.
5) or in a spherical section of half thickness $\theta \simgt 60^\circ$.
The second (radial) component is probably made up of a thin shell of matter
being pushed away in a jet--like configuration, where the emissivity
decreases from the inner to the outer edge of the shell. Again, the best
agreement with the observations is obtained in the case where the effect of
anisotropic line emission is included. The best fit is obtained with a
shell of thickness t $< 3 R_{min}$, in a cone of half--aperture $\Theta_0
\approx 12^\circ$ and seen at $i = 0^\circ$. The emissivity is represented
by a gaussian peaked at $r= R_{min}$, with 2$\sigma^2$ = 0.181 R$_{min}^2$,
and where the number of clouds decreases as $\propto r^{-5}$. We obtain
however model profiles with blue wings and roughly flat tops for both
power--law and gaussian emissivity, with $\Theta_0 = 12^\circ$, in some
cases with $i$\ up to $\sim 30^\circ$.  A very weak blue hump (due to the
approaching gas in the double stream) should be seen on the blue wing of
H\b\ if anisotropy is the only thing taken into account. This would
correspond to a $4 \sigma$\ feature which could  be suppressed by a small
amount of extinction due to dust on the non illuminated face of the clouds
or even be lost in the aborption features contaminating the blue wing of
H\b.

We note that it  is also possible to roughly reproduce the observed profiles
assuming that gas is infalling and decelerated by radiation pressure.
However, in light of the analysis of the BLR dynamics outlined above,
infall is not favored by the variations of the line.


\section{Conclusions}

The most likely explanation for the complex Balmer line profiles of OQ 208
involves outflow of the emitting gas probably radiatively accelerated in a
biconical geometry, plus emission from an ensemble of clouds which might be
spherically symmetric or confined in a thick cylinder.
Emission from infalling gas and/or from a rotating disk (heavily obscured)
is not favored by our analysis, but cannot be ruled out and remains a
competing alternative. Optical depth effects in the Balmer lines must be
taken into account to properly understand the asymmetries observed in the
line profiles. Several important questions follow  from the present
investigation. The most important are related to the peculiarity in the
FeII emission, and to the existence of an obscured or self absorbed line
component.

Observations with Hubble Space Telescope would allow us to measure radial
velocities for the strongest high ionization lines that are located in the
ultraviolet. They would clarify whether there is also a systematic shift
between LIL and HIL for Seyfert galaxies. Some predictions are possible on
the basis of the model outlined in the previous section. If the red peak of
the Balmer lines arises in an outflowing component we do not expect to see
a redshift difference between the HIL and LIL. If the LIL arise from
infalling clouds we expect to see a large velocity difference between the
HIL and LIL. In the latter case the HIL and LIL would arise from opposite
sides of the BLR and from opposite faces of the clouds. A UV spectrum for
OQ208 would therefore provide an unambiguous test of our model.

Spectroscopic monitoring of OQ 208 would lead to other valuable
information. Temporal sampling in the interval from one to a few months
would constrain the line response to continuum changes. We observed
variations in the red hump as well as in the blue wing of H\b. Adequate
monitoring would allow to determine whether such changes occur
simultaneously and if not, which occurred first.
The strong
variations observed in the line profiles of OQ208 (occurring on a rather
long timescale) suggest that monitoring holds promise of a large return
from relatively few observations covering a period of few years.
Although OQ 208 is relatively faint, with present instrumentation it is
possible to obtain high S/N spectra (as demonstrated by the Kitt Peak
spectrum employed in this study). OQ208 may be one of the first AGN for
which an unambiguous model of the BLR is possible.

\section{Acknowledgements}

PM would like to acknowledge the financial support provided by Italian
Consiglio Nazionale delle Ricerche, which allowed him to spend a short
post-doc period at the University of Alabama, where this work was almost
completely done. R. Cohen and W. Zheng kindly provided us with the spectrum
taken at Lick Observatory in Apr. 1985.

Extragalactic astronomy at the University of Alabama is supported under
EPSCoR grant RII--8996152. The INT is operated in the island of La Palma by
the Royal Greenwich Observatory in the Spanish Observatorio del Roque de
los Muchachos del Instituto de Astrofisica de Canarias.

\newpage
\section{References}

\REF

\REF
Appenzeller I., Ostreicher R., 1988, AJ 95, 45
\REF
Blake G. M., Argue A. N., Kenworthy C. M., 1970, ApL 6, 167
\REF
Blumenthal G. R., Mathews W. G., 1975, ApJ 198, 517
\REF
Burbidge G. M.,  Strittmatter P.A., 1972, ApJ 172, L37
\REF
Charlot P., 1990, A\&A 229, 51
\REF
Chen K., Halpern J. P., 1989, ApJ 344, 115
\REF
Chen K., Halpern J. P., Filippenko A. V., 1989 ApJ  339, 742
\REF
Crenshaw D. M., Peterson B. M., 1986, PASP 98, 185
\REF
Ferland G. J., Peterson B. M., Horne K., Welsh W. F., Nahar S. N.,  1992,
ApJ 387, 95
\REF
Gaskell C. M., 1983a, ApJL 267, L1
\REF
Gaskell C. M., 1983b, in Proceedings of the Liege Conference on Quasars \&
\ Gravitational Lenses, (Liege: Institut d' Astrophysique), 473
\REF
Halpern J. P., 1990, ApJ 365, L51
\REF
Halpern J. P., Filippenko A. V., 1988, Nature 331, 46
\REF
Heckman T., Miley G. K., Green R .F., 1984, ApJ 281, 525
\REF
Jackson N., Browne I. W. A., 1991, MNRAS 250, 414
\REF
Joly M., 1991, A \&\ A 242, 49
\REF
\REF
Marziani P., 1991, PhD Thesis, (Trieste: SISSA)
\REF
Marziani P., Calvani M.,\&\ Sulentic J. W., 1992, ApJ 393, 658
\REF
Mathews W. G., 1982, ApJ  258, 425
\REF
Mathews W. G., \&\ Capriotti E.R., 1986, in Astrophysics of Active Galaxies and
Quasi Stellar Objects (Mill Valley: University Science Books) J.S. Miller
(Ed.), p. 185
\REF
Mathews W. G., \&\ Veilleux S., 1989, ApJ 336, 93
\REF
Mathews W. G., \&\ Ferland G. J., 1987, ApJ 323, 456
\REF
Miller J. S., \&\ Peterson B. M., 1990, ApJ 361, 91
\REF
Morris S. L., \&\ Ward M. J., 1988, ApJ 340, 713
\REF
Netzer H., 1990, in Active Galactic Nuclei, p. 57 (Berlin:Springer)
\REF
Orr  M. J. L., \&\ Browne I. W. A., 1982, MNRAS 200, 1067
\REF
Osterbrock D. E., 1978, PhysScripta 17, 137
\REF
Osterbrock D. E., 1991, RepProgresPhys 54, 579
\REF
Osterbrock D. E., \&\ Cohen R., 1979, MNRAS 187, 61p (OC79)
\REF
Padovani P., Rafanelli P., 1988, A\&A 205, 53
\REF
Penston M., 1991, in Variability of Active Galactic Nuclei
Ed. H. R. Miller (Cambridge: Cambrige University Press), 343
\REF
Perez E., 1987, Ph.D. Thesis, University of Sussex
\REF
Perez E., Penston M. V., Tadhunter C., Mediavilla E., Moles M., 1988, MNRAS
230, 353
\REF
Peterson B. M., Ferland G. J., 1986,  Nature 324, 345
\REF
Phillips M. M., 1978, ApJS 38, 187
\REF
Rees M. J., Netzer H., Ferland G. J., 1989, ApJ 347, 640
\REF
Robinson A., Perez E., Binette L., 1990, MNRAS 246, 349
\REF
Ryle M.,  Poley G.G., 1970 ApL 4, 137
\REF
Rafanelli P., Marziani P., 1992, AJ 103, 743
\REF
Stockton A., Farnham J., 1991, ApJ 371, 525
\REF
Sulentic J.W.,  1989, ApJ 343, 54
\REF
Sulentic J. W., Calvani M., Marziani P., Zheng W., 1990, ApJ 355, L15
\REF
Sulentic, J. W., Zheng, W. and Arp, H., 1990, PASP 102, 1275.
\REF
Ulrich M.-H., Altamore A., Boksenberg A., Bromage G., Clavel J., Elvius A.,
Penston M. V., 1985, Nature 313, 747
\REF
Veilleux S., Zheng W., 1991, ApJ, 377, 89
\REF
Wallerstein G., Wilson L.A., Salzer J., Brugel E., 1984, A\&A 133, 137
\REF
Waltman E. B., Fiedler R. L., Johnston K. J., Spencer J. H., Florkowski D. R.,
Jostie F. J., Mc Carthy D. D., Matsakis D. N., 1991, ApJS 77, 379
\REF
Wills B. J., Browne I. W. A., 1986, ApJ 302, 56
\REF
Wills B. J., Wills D., 1986, IAU Symp. 119, G. Swarup, V.K. Kapahi (Eds.),
p. 215
\REF
Zheng W., 1991, ApJ 382, L55
\REF
Zheng W., Binette L., Sulentic J. W., 1990, ApJ 365, 115
\REF
Zheng W., Veilleux S., Grandi S. A., 1991, ApJ 381, 418
\newpage
\section{Figure Captions}

\REF
Fig. 1: The spectrum of OQ 208 obtained on 1989 July 31. Horizontal scale
is wavelength in \AA, vertical scale is flux in units of 10$^{-15}$\ erg
cm$^{-2}$ s$^{-1}$.

\REF
Fig. 2: H\b\ and H\a\ profile of OQ208 at different epochs (a) and (d) 1985
Jun. 15, (b) and (e) 1989 Jul. 31, (c) 1991 Feb. 19. and (f) 1990 Apr. 04.
All spectra have been normalized to the same [OIII]\l\l 4959,5007 flux,
with the exception of the spectrum taken on 1990 Apr. 4 has been
arbitrarily scaled to roughly match the continuum level of the 1991 Feb. 19
spectrum. Units are as for Fig. 1. Vertical and horizontal scales have been
chosen in order to evidentiate the variations in the continuum as well as
in the line profiles. Note that B--band absorption heavily contaminates the
blue wing of H\a\ of the profiles in panels (b) and (d). Correction for
B--band absorption has been applied only to the profile shown in panel (f).

\REF
Fig. 3: H\b\ profile difference between the spectrum taken on 1991 Feb. 19
and the weighted average of the spectra obtained in 1985. Units are as for
Fig. 1.

\REF
Fig. 4: Correlations between H\b\ and continuum fluxes and shift. Vertical
scale is radial velocity difference (in km/s) between the centroid of the
red hump of H\b\ broad component (see text) and peaks of the narrow
component of H\b. (a) Horizontal scale is ratio flux of total H\b$_{BC}$\
flux and H\b$_{NC}$\ flux; (b) horizontal scale is approximately flux
within $\pm$ 13 \% HWZI of H\b$_{BC}$\ normalized by the H\b$_{NC}$\ flux;
(c) horizontal scale is specific flux of the continuum at 4800 \AA. Open
circles: centroid defined by eye estimate; filled circles: centroid defined
as weighted average of wavelength over the third power of the intensity.
Error bars refer to 1$\sigma$\ level of uncertainty.


\REF
Fig. 5: Best fit of the H\b\ profile observed in Feb. 1991. Dotted line:
fit to the red hump + line base. Dashed line: fit to the line base. The
straight dot--dashed line indicates the position of the narrow component of
H\b, which has been removed.

\end{flushleft}

\end{document}